\documentclass[sigconf,screen]{acmart}
\pdfoutput=1
\usepackage[english]{babel}
\usepackage[T1]{fontenc}
\usepackage[utf8]{inputenc}

\usepackage{algpseudocode}
\usepackage{amsfonts}
\usepackage{amsmath}
\usepackage{amssymb}
\usepackage{amsthm}
\usepackage{balance}
\usepackage{bm}
\usepackage{booktabs}
\usepackage{comment}
\usepackage[inline]{enumitem}
\usepackage{mathtools}
\usepackage{pgfplots}
\usepackage{xparse}
\usepackage{xspace}
\usepackage{xstring}
\usepackage{multirow}
\usepackage{makecell}
\usepackage[separate-uncertainty]{siunitx}

\pgfplotsset{compat=1.14}
\usetikzlibrary{arrows.meta}

\DeclarePairedDelimiter{\ceil}{\lceil}{\rceil}
\DeclarePairedDelimiter{\floor}{\lfloor}{\rfloor}

\NewDocumentCommand{\transp}{}{%
  \top%
}

\NewDocumentCommand{\mat}{m}{%
  \ensuremath{\bm{#1}}\xspace%
}

\NewDocumentCommand{\VEC}{o o m}{%
  \IfValueTF{#1}{%
    \IfValueTF{#2}{%
      \IfStrEq{#1}{}{%
        \ensuremath{#3_{I_{#2}}}\xspace%
      }{%
        \ensuremath{#3_{I_{#2}}^{(#1)}}\xspace%
      }%
    }{%
      \IfStrEq{#1}{}{%
        \ensuremath{#3}\xspace%
      }{%
        \ensuremath{#3^{(#1)}}\xspace%
      }%
    }%
  }{%
    \ensuremath{#3}\xspace%
  }%
}

\RenewDocumentCommand{\vec}{o o m}{%
  \VEC[#1][#2]{\bm{#3}}%
}

\RenewDocumentCommand{\O}{m}{%
  \ensuremath{\mathcal{O}(#1)}\xspace%
}

\NewDocumentCommand{\N}{o o}{%
  \IfValueTF{#1}{%
    \IfValueTF{#2}{%
      \ensuremath{\mathbb{N}^{#1 \times #2}}\xspace%
    }{%
      \ensuremath{\mathbb{N}^{#1}}\xspace%
    }%
  }{%
    \ensuremath{\mathbb{N}}\xspace%
  }%
}

\NewDocumentCommand{\R}{o o}{%
  \IfValueTF{#1}{%
    \IfValueTF{#2}{%
      \ensuremath{\mathbb{R}^{#1 \times #2}}\xspace%
    }{%
      \ensuremath{\mathbb{R}^{#1}}\xspace%
    }%
  }{%
    \ensuremath{\mathbb{R}}\xspace%
  }%
}

\NewDocumentCommand{\A}{}{%
  \mat{A}%
}

\NewDocumentCommand{\matL}{}{%
  \mat{L}%
}

\NewDocumentCommand{\LT}{}{%
  \ensuremath{\matL^{\transp}}%
}

\NewDocumentCommand{\M}{}{%
  \mat{M}%
}

\NewDocumentCommand{\Minv}{}{%
  \ensuremath{\M^{-1}}\xspace%
}

\NewDocumentCommand{\matP}{}{%
  \mat{P}%
}

\NewDocumentCommand{\B}{o}{%
  \vec[][#1]{b}%
}

\NewDocumentCommand{\x}{o o}{%
  \vec[#1][#2]{x}%
}

\NewDocumentCommand{\res}{o o}{%
  \vec[#1][#2]{r}%
}

\NewDocumentCommand{\dir}{o o}{%
  \vec[#1][#2]{p}%
}

\NewDocumentCommand{\z}{o o}{%
  \vec[#1][#2]{z}%
}

\NewDocumentCommand{\I}{o}{%
  \IfValueTF{#1}{%
    \ensuremath{I_{#1}}\xspace%
  }{%
    \ensuremath{I}\xspace%
  }%
}

\NewDocumentCommand{\f}{}{%
  \ensuremath{f}\xspace%
}

\NewDocumentCommand{\nredu}{}{%
  \ensuremath{\phi}\xspace%
}

\NewDocumentCommand{\nfail}{}{%
  \ensuremath{\psi}\xspace%
}

\NewDocumentCommand{\nn}{}{%
  \ensuremath{N}\xspace%
}

\NewDocumentCommand{\npn}{}{%
  \ensuremath{m}\xspace%
}

\NewDocumentCommand{\np}{}{%
  \ensuremath{M}\xspace%
}

\NewDocumentCommand{\latency}{}{%
  \ensuremath{\lambda}\xspace%
}

\NewDocumentCommand{\invbw}{}{%
  \ensuremath{\mu}\xspace%
}

\NewDocumentCommand{\CommOverhead}{}{%
  \ensuremath{\mathcal{O}}\xspace%
}

\NewDocumentCommand{\parabolicfem}{}{%
M1\xspace%
}

\NewDocumentCommand{\offshore}{}{%
M2\xspace%
}

\NewDocumentCommand{\gthree}{}{%
M3\xspace%
}

\NewDocumentCommand{\thermal}{}{%
M4\xspace%
}

\NewDocumentCommand{\emilia}{}{%
M5\xspace%
}

\NewDocumentCommand{\geo}{}{%
M6\xspace%
}

\NewDocumentCommand{\serena}{}{%
M7\xspace%
}

\NewDocumentCommand{\audikw}{}{%
M8\xspace%
}

\NewDocumentCommand{\relresdiff}{m}{%
  \ensuremath{\Delta_{#1}}\xspace%
}

\allowdisplaybreaks

\newfloat{algorithm}{t}{lop}
\floatname{algorithm}{Algorithm}


\copyrightyear{2019}
\acmYear{2019}
\setcopyright{acmlicensed}
\acmConference[ICPP 2019]{48th International Conference on Parallel Processing}{August 5--8, 2019}{Kyoto, Japan}
\acmPrice{15.00}
\acmDOI{10.1145/3337821.3337849}
\acmISBN{978-1-4503-6295-5/19/08}

\begin{document}

\title[How to Make the PCG Method Resilient Against Multiple Node Failures]
      {How to Make the Preconditioned Conjugate Gradient Method Resilient
       Against Multiple Node Failures}

\author{Carlos Pachajoa}
\orcid{0000-0002-0688-8130}
\affiliation{%
\institution{University of Vienna}
\department{Faculty of Computer Science}
\city{Vienna}
\country{Austria}}
\email{carlos.pachajoa@univie.ac.at}

\author{Markus Levonyak}
\orcid{0000-0002-5131-6318}
\affiliation{%
\institution{University of Vienna}
\department{Faculty of Computer Science}
\city{Vienna}
\country{Austria}}
\email{markus.levonyak@univie.ac.at}

\author{Wilfried N. Gansterer}
\authornote{Corresponding author}
\affiliation{%
\institution{University of Vienna}
\department{Faculty of Computer Science}
\city{Vienna}
\country{Austria}}
\email{wilfried.gansterer@univie.ac.at}

\author{Jesper Larsson Träff}
\affiliation{%
\institution{TU Wien}
\department{Faculty of Informatics}
\city{Vienna}
\country{Austria}}
\email{traff@par.tuwien.ac.at}

\renewcommand*{\shortauthors}{C. Pachajoa, M. Levonyak, W.N. Gansterer, and
J.L. Träff}

\begin{abstract}
We study algorithmic approaches for recovering from the failure of several
compute nodes in the parallel preconditioned conjugate gradient (PCG) solver on
large-scale parallel computers.
In particular, we analyze and extend an exact state reconstruction (ESR)
approach, which is based on a method proposed by Chen (2011).
In the ESR approach, the solver keeps redundant information from previous
search directions, so that the solver state can be fully reconstructed if a
node fails unexpectedly.
ESR does not require checkpointing or external storage for saving dynamic
solver data and has low overhead compared to the failure-free situation.

In this paper, we improve the fault tolerance of the PCG algorithm based on the
ESR approach.
In particular, we support recovery from simultaneous or overlapping failures of
several nodes for general sparsity patterns of the system matrix, which cannot
be handled by Chen's method.
For this purpose, we refine the strategy for how to store redundant information
across nodes.
We analyze and implement our new method and perform numerical experiments with
large sparse matrices from real-world applications on 128 nodes of the Vienna
Scientific Cluster (VSC).
For recovering from three simultaneous node failures we observe average runtime
overheads between only 2.8\% and 55.0\%.
The overhead of the improved resilience depends on the sparsity pattern of the
system matrix.

\end{abstract}

\maketitle


\section{Introduction}
\label{sec:introduction}

One of the major challenges in current and, even more, in future
high-performance computing (HPC) is the increasing failure rate caused by
increasing complexity and an ever-growing number of interconnected components
\cite{snir2014, gupta2017fls}.
Among the most common types of failures in large-scale parallel computers are
\emph{fail-stop failures}, where a failing process stops and its data is lost
\cite{Schlichting1983a, Elnozahy2002a, Schroeder2010a, chen2011esr}.
In order to support long-running scientific applications also on failure-prone
HPC systems, new strategies and resilient algorithms are necessary
\cite{Herault2015a}.

We consider the \emph{preconditioned conjugate gradient} (PCG) method
(cf.\ Sec.~\ref{sec:background:pcg}), an important iterative method for solving
symmetric and positive-definite (SPD) sparse linear systems on large-scale
parallel computers.
We study settings where the PCG solver is executed on parallel computers that
are susceptible to \emph{node failures}, a common type of fail-stop failures
which is critical in practice \cite{Herault2015a}.
In contrast to previous work that also aims to make the PCG method resilient
against unexpected node failures without expensive checkpointing
(cf.\ Sec.~\ref{sec:introduction:related}), we do not only consider single node
failures but multiple node failures which occur simultaneously or are
overlapping in time.
The \emph{exact state reconstruction} (ESR) approach
(cf.\ Sec.~\ref{sec:background:esr}), which was introduced by Chen
\cite{chen2011esr} and refined by Pachajoa et al.\ \cite{Pachajoa2018a}, was
shown to be the most efficient checkpointing-free algorithmic fault-tolerance
technique for protecting the PCG solver against unexpected single node failures
\cite{Pachajoa2018a}.

In this paper, we assume that a large sparse linear system \(\A \x = \B\),
where \A is SPD, is given and shall be solved with the PCG method on a parallel
computer.
The preconditioner \M for solving this linear system is either explicitly or
implicitly given.
We enhance the ESR approach so that it becomes capable of protecting the PCG
method against unexpected simultaneous or overlapping failures of multiple
nodes, theoretically analyze the communication overhead for this improved
resilience, and demonstrate low runtime overheads in numerical experiments.
Although we cannot provide details due to space restrictions, our proposed
algorithmic modifications can also be applied to the ESR approach
\cite{chen2011esr} for the \emph{Jacobi}, \emph{Gauss-Seidel}, \emph{successive
overrelaxation} (SOR), \emph{symmetric successive overrelaxation} (SSOR),
\emph{split preconditioner conjugate gradient} (SPCG) \cite{Pachajoa2018a} and
\emph{preconditioned bi-conjugate gradient stabilized} (BiCGSTAB) algorithms in
order to make them resilient against multiple simultaneous or overlapping node
failures.

\subsection{Problem setting and assumptions}
\label{sec:introduction:setting}

We consider the parallel execution of the PCG solver on \nn compute nodes of a
distributed-memory parallel computer, which communicate over an interconnection
network.
Each compute node consists of \npn processors such that the solver is executed
on \(\np \coloneqq \nn \times \npn\) processors in total.
Each processor shares its memory with all other processors on the same
compute node.

In the event of a \emph{single processor failure}, exactly one processor on one
compute node fails, but the shared memory on this node stays intact.
Hence, no data is lost and the remaining processors on the node can take over
the workload of the failed processor.
The same holds for \emph{multiple processor failures} where at least one
processor per compute node survives.
In this paper, we consider a more complicated event: a \emph{node failure},
where a compute node fails as a whole and the data in the memory of the
affected node is lost.

Node failures may occur for several reasons: for example, all \npn processors
of a node fail, the shared memory of a node gets corrupted, or a node loses its
connection to the interconnection network.
In case of a \emph{single node failure}, exactly one compute node fails
at a time.
If more than one node fails at a time, we talk about \emph{multiple node
failures}.
Nodes that continue working after a node failure and keep all their data are
called \emph{surviving nodes}.
A node that becomes unavailable after a node failure is referred to as a
\emph{failed node}, and a node that takes the place of a failed node in the
recovery process is called a \emph{replacement node} (which could be a spare
node or one of the surviving nodes).

\subsubsection{Failure detection and node replacement}
\label{sec:introduction:setting:detection}

We assume the availability of a parallel runtime environment that provides
functionality comparable to state-of-the-art implementations of the
industry-standard \emph{Message Passing Interface} (MPI) \cite{MPIF2015a}.
This particularly comprises efficient collective communication capabilities
like the \emph{Allreduce} function of MPI.
Beyond that, we assume that the underlying runtime environment supports some
basic fault-tolerance features.
A prototypical example is the MPI extension \emph{User Level Failure
Mitigation} (ULFM) \cite{Bland2013a, MPIF2017a} which supports the detection of
node failures, the prevention of indefinitely blocking synchronizations or
communications, the notification of the surviving nodes that a failure has
occurred and which nodes have failed, and a mechanism for providing replacement
nodes.

\subsubsection{Data distribution}

Analogously to \cite{chen2011esr}, we assume that the fundamental
problem-defining static input data, i.e., the system matrix \(\A \in
\R[n][n]\), the right-hand-side vector \(\B \in \R[n]\), and the preconditioner
\(\M \in \R[n][n]\) (cf.\ Sec.~\ref{sec:background:pcg}), can be retrieved from
reliable external storage (e.g., from a checkpoint prior to entering the linear
solver, cf.\ the \emph{ABFT\&PeriodicCkpt} algorithm \cite{Bosilca2014a,
Bosilca2015a, Herault2015a}, see Sec.~\ref{sec:introduction:related}),
and thus does not have to be reconstructed after a node failure.

In accordance with typical distributions of sparse matrices in widely used
high-performance numerical libraries like PETSc \cite{Balay1997a}, we consider
a block-row data distribution of all matrices and vectors among the \nn nodes
of the parallel computer.
More precisely, every node owns blocks of \(n / \nn\) contiguous rows (if
\(n = c \nn\) with \(c \in \mathbb{N}\), otherwise some nodes own
\(\floor{n / \nn}\) and others \(\ceil{n / \nn}\) rows) of both the matrices
\A and \M as well as each of the vectors \B, \x, and all other vectors
maintained by the PCG solver (cf. Sec.~\ref{sec:background:pcg}).
On one node, the owned block rows, which are stored in the shared memory of the
node, are evenly distributed among the \npn processors, i.e., each processor
owns (approximately) \(n / \np\) rows of each of the matrices and vectors.
Globally used scalars are replicated on all \nn nodes.

Since each node owns block rows of all matrices and vectors, a node failure
affects a part of each matrix and vector.
Block rows of dynamic data which were owned by the failed node are lost and
need to be reconstructed on the replacement node
(cf.\ Sec.~\ref{sec:background:esr}).

\subsubsection{Notation}

We use a notation similar to \cite{Agullo2016a} to denote sections of matrices
and vectors.
We refer to the set of all indices as \(\I \coloneqq \{1, 2, \dots, n\}\).
The cardinality \(n\) of \I is equal to the size of the vectors.
The index subset representing the rows assigned to node \(i\) is denoted as
\I[i].
Given a vector \vec[j]{v}, where \(j\) denotes the iteration number of the
linear solver, \vec[j][i]{v} refers to the subset of elements of the vector at
iteration \(j\) owned by node \(i\).
Row and column selections of a matrix \mat{B} are designated with index sets as
well: \(\mat{B}_{\I[i], \I[k]}\) refers to the selection of rows and columns of
\mat{B} corresponding to the index subsets \I[i] and \I[k], respectively.

The failed node is referred to as node \f, and its index set is hence denoted
as \I[\f].
With the \emph{state} of an iterative solver we mean the---not necessarily
minimal---set of data that completely defines the future behavior of this
iterative solver.
The state of the PCG solver in iteration \(j\) can be defined as comprising the
iterate \x[j] (i.e., the current approximation to the solution \x), the
residual \res[j], the preconditioned residual \z[j], and the search direction
\dir[j].

\subsection{Related work}
\label{sec:introduction:related}

The existing literature on resilient iterative linear solvers distinguishes
between two different kinds of failures: soft errors and node failures.
The former refers to spontaneous changes of the state of the solver (e.g., bit
flips), potentially leading to a wrong result.
In this category, we find work by Sao and Vuduc \cite{sao2013} which proposes
strategies to ensure that the \emph{conjugate gradient} (CG) method will
converge to the right solution after a soft error.
However, their approach requires some operations to be performed reliably.

Bronevetsky and de Supinski \cite{Bronevetsky2008a} evaluate the effects of
soft errors on iterative linear solvers including the CG method.
Bit flips are introduced at random times and positions, and the effects are
classified according to the resulting runtimes and solution errors.
Dichev and Nikolopoulos \cite{Dichev2016a} propose and experimentally evaluate
a specific form of \emph{dual modular redundancy}, where \emph{all}
computations are performed twice for improved redundancy, in order to detect
and correct soft errors in the PCG method.
More work in the area of soft error detection and correction in the CG and PCG
methods has been published by Shantharam et al. \cite{Shantharam2012a} and Fasi
et al. \cite{Fasi2016a}.
All these approaches have in common that they are not applicable to our problem
of the PCG solver subject to node failures.

Pachajoa and Gansterer \cite{pachajoa2018cgmg} experimentally evaluate the
inherent resilience of the CG method after a single node failure.
The currently in practice most commonly used class of fault-tolerance
techniques to cope with node failures is \emph{checkpoint/restart} (\emph{C/R}).
These techniques frequently save the current state of a running application and
roll back to the latest saved state in case of a node failure.
C/R has been investigated as a general-purpose technique (e.g. \cite{tiwari2014lazy}),
and also for specific problem settings (e.g. \cite{ltaief2008ftaheat}).
Herault et al. \cite{Herault2015a} provide a comprehensive overview of
different variants of this approach.

To avoid the overhead of continuously saving the state, Chen \cite{chen2011esr}
exploits specific properties of the PCG solver and other iterative methods such
that the state of the solver can be recovered without checkpointing after a
single node failure occurred.
For storing the necessary redundant information, he chooses the intuitive
approach of sending it to the closest node (cf.\ Sec.~\ref{sec:single}).
Pachajoa et al. \cite{Pachajoa2018a} refine Chen's approach \cite{chen2011esr}
by distinguishing different common types of preconditioners.

In \cite{Pachajoa2018a}, the approach from \cite{chen2011esr} is experimentally
compared to a heuristic strategy proposed by Langou et
al.\ \cite{langou2007recovery}.
This heuristic \emph{interpolation/restart} strategy is applicable to iterative
linear solvers in general to recover from a node failure by approximating the
lost iterate.
The submatrix of the replacement node is used to produce an interpolated
approximation of the iterate before the node failure occurred.
Agullo et al. \cite{Agullo2016a, agullo2013resilientkrylov} extend this
approach by using all the information of the matrix in the interpolation, which
produces a reconstructed iterate whose error norm is guaranteed to be smaller
than the error norm of the iterate before the node failure, albeit with
significant communication overhead.

Bosilca et al. \cite{Bosilca2014a, Bosilca2015a, Herault2015a} introduce the
\emph{ABFT\&PeriodicCkpt} algorithm, which combines \emph{algorithm-based fault
tolerance} (ABFT) with periodic checkpointing in order to make entire
applications (and not just operations that can be protected by ABFT) resilient
to node failures.
The longer the phases protected by ABFT, the fewer checkpoints are necessary
and the cheaper resilience against node failures usually becomes (assuming that
the used ABFT method is more efficient than checkpointing).

\subsection{Contributions of this work}
\label{sec:introduction:contributions}

To the best of our knowledge, there has been neither a detailed discussion nor
a thorough analysis of a generalized ESR approach for protecting the PCG method
against \emph{multiple} node failures.
In this paper, we propose a strategy that precisely defines how and where to
store redundant information so that the PCG solver becomes resilient against
multiple node failures.
In particular, we analyze the communication overhead and evaluate the
performance penalty for the improved resilience capabilities of being able to
tolerate multiple simultaneous node failures.
In numerical experiments on the \emph{Vienna Scientific Cluster} (VSC), a
medium-scale HPC system, we eventually demonstrate the low runtime overhead of
our new strategies for the improved resilience.

\medskip

\noindent The remainder of this paper is organized as follows.
In Sec.~\ref{sec:background}, we briefly review the PCG method and the ESR
approach as presented in \cite{chen2011esr, Pachajoa2018a}.
Afterwards, in Sec.~\ref{sec:single}, we summarize the concept of
Chen~\cite{chen2011esr} for keeping redundant information in the ESR approach
such that single node failures can be tolerated.
Then, in Sec.~\ref{sec:multiple}, we propose modifications and extensions of
the ESR approach for systems that are prone to multiple simultaneous or
overlapping node failures.
Apart from that, we theoretically analyze the communication overhead of our
novel algorithm for supporting multiple node failures.
In Sec.~\ref{sec:sparsity}, we discuss the most important aspects regarding the
impact of the sparsity pattern of the system matrix.
Next, in Sec.~\ref{sec:implementation}, we summarize our implementation for
conducting numerical experiments.
Subsequently, in Sec.~\ref{sec:experiments}, we present our experimental
results and discuss how they relate to our theoretical analysis.
Finally, our conclusions are summarized in Sec.~\ref{sec:conclusions}.

\section{Algorithmic background}
\label{sec:background}

In this section, we first review the PCG method in
Sec.~\ref{sec:background:pcg} and afterwards the ESR approach
\cite{chen2011esr, Pachajoa2018a} in Sec.~\ref{sec:background:esr}.
In Sec.~\ref{sec:single}, we then look at the details of how Chen
\cite{chen2011esr} is keeping redundant information in order to guarantee
protection against single node failures.

\begin{algorithm}
{\fontsize{9}{15}\selectfont
\begin{algorithmic}[1]
  \State \(\res[0] \coloneqq \B - \A \x[0], \z[0] \coloneqq \Minv \res[0],
    \dir[0] \coloneqq \z[0]\)
  \For {\(j = 0, 1, \dots, \text{ until convergence}\)}
  \State \(\alpha^{(j)} \coloneqq \res^{(j) \transp} \z[j] /
    \dir^{(j) \transp} \A \dir[j]\)
    \label{alg:pcg:alpha}
  \State \(\x[j+1] \coloneqq \x[j] + \alpha^{(j)} \dir[j]\)
    \label{alg:pcg:x}
  \State \(\res[j+1] \coloneqq \res[j] - \alpha^{(j)} \A \dir[j]\)
    \label{alg:pcg:r}
  \State \(\z[j+1] \coloneqq \Minv \res[j+1]\)
    \label{alg:pcg:z}
  \State \(\beta^{(j)} \coloneqq \res^{(j+1) \transp} \z[j+1] /
    \res^{(j) \transp} \z[j]\)
    \label{alg:pcg:beta}
  \State \(\dir[j+1] \coloneqq \z[j+1] + \beta^{(j)} \dir[j]\)
    \label{alg:pcg:p}
  \EndFor
\end{algorithmic}
}
\caption{Preconditioned conjugate gradient (PCG) method
  \cite[Alg.~9.1]{saad2003iterative}}
\label{alg:pcg}
\end{algorithm}

\begin{algorithm}
{\fontsize{9}{18}\selectfont
\begin{algorithmic}[1]
\State Retrieve the static data \(\A_{\I[\f], \I}\), \(\matP_{\I[\f], \I}\),
  and \B[\f]
\State Gather \(\res[j]_{\I \setminus \I[\f]}\) and
  \(\x[j]_{\I \setminus \I[\f]}\)
\State Retrieve the redundant copies of \(\beta^{(j-1)}\), \dir[j-1][\f], and
  \dir[j][\f] \label{alg:esr:pcg:2:retrieve}
\State Compute \(\z[j][\f] \coloneqq \dir[j][\f] - \beta^{(j-1)} \dir[j-1][\f]\)
\State Compute \(\vec{v} \coloneqq \z[j][\f] -
  \matP_{\I[\f], \I \setminus \I[\f]} \res_{\I \setminus \I[\f]}^{(j)}\)
  \label{alg:esr:pcg:2:pc_rhs}
\State Solve \(\matP_{\I[\f], \I[\f]} \res[j][\f] = \vec{v}\) for \res[j][\f]
  \label{alg:esr:pcg:2:pc_solve}
\State Compute \(\vec{w} \coloneqq \B[\f] - \res[j][\f] -
  \A_{\I[\f], \I \setminus \I[\f]} \x[j]_{\I \setminus \I[\f]}\)
  \label{alg:esr:pcg:2:rhs}
\State Solve \(\A_{\I[\f], \I[\f]} \x[j][\f] = \vec{w}\) for \x[j][\f]
  \label{alg:esr:pcg:2:solve}
\end{algorithmic}
}
\caption{ESR reconstruction phase for the PCG method on the replacement node \f
  (\(\matP \coloneqq \Minv\) is given) \cite[Alg.~4]{Pachajoa2018a}}
\label{alg:esr:pcg:2}
\end{algorithm}

\subsection{Preconditioned conjugate gradient method}
\label{sec:background:pcg}

The (P)CG method iteratively solves a linear system \(\A \x = \B\),
where both the SPD matrix \(\A \in \R[n][n]\) and the right-hand-side vector
\(\B \in \R[n]\) are given.
The search directions \dir[j], along which the quadratic potential defined by
\A is minimized, are chosen to be \A-orthogonal, i.e., \(\dir^{(j) \transp} \A
\dir[k] = 0\) for all \(j \neq k\).
The residual \res[j] is defined as \(\res[j] = \B - \A \x[j]\).

In the PCG method, which is listed in Alg.~\ref{alg:pcg}, a preconditioner
\(\M \in \R[n][n]\) is used for accelerating convergence.
Instead of the original system, the linear system \(\Minv \A \x = \Minv \B\)
is solved.
\M is assumed to be an SPD matrix as well \cite[p. 276]{saad2003iterative} and
is to be chosen such that \(\kappa(\Minv \A) < \kappa(\A)\), where
\(\kappa(\mat{B})\) denotes the condition number of a matrix \mat{B}.
The PCG method stores the distributed vectors \x[j], \res[j], \z[j], \dir[j],
and \(\A \dir[j]\) as well as the replicated scalars \(\alpha^{(j)}\) and
\(\beta^{(j)}\), which in total require memory for \(5n + 2N\) floating-point
numbers (not including the static data \A, \M, and \B).

\subsection{Exact state reconstruction (ESR)}
\label{sec:background:esr}

In contrast to checkpoint/restart methods, which---even in the failure-free
case---impose a usually considerable runtime overhead due to continuously
saving the state of the solver \cite{Herault2015a}, the ESR approach
\cite{chen2011esr, Pachajoa2018a} is able to exploit the algorithmic properties
of the PCG solver so that the complete state can be reconstructed after a node
failure with low overhead.
During the failure-free PCG iterations, only little or---depending on the
sparsity structure of \A---no additional communication is necessary for
achieving this.
Only the local memory requirements are slightly higher than in the standard
(non-resilient) PCG variant.

At iteration $j$ of the PCG algorithm, the product \(\mat{A} \dir[j]\) is
computed in a sparse matrix-vector multiplication (SpMV) operation
(cf.\ Alg.~\ref{alg:pcg}, lines~\ref{alg:pcg:alpha} and~\ref{alg:pcg:r}).
In the non-resilient PCG solver, all but the own block \dir[j][i] can be
dropped on node \(i\) after the product has been computed.
In the resilient algorithm, we also drop most of \dir[j] on each node after the
SpMV.
The crucial difference to non-resilient PCG is that now each node also stores
the elements of at least one other node in addition to its own block (for
details see Sec.~\ref{sec:single}).

Overall, there is a redundant copy of each element of \dir[j] after computing
\(\mat{A} \dir[j]\).
Hence, in case of a node failure, this redundant copy can be sent to the
replacement node in order to completely recover the most recent search
direction.
For the reconstruction of the complete state after a node failure, we need to
recover the two most recent search directions \cite[Sec.~5.2]{chen2011esr} and
thus have to keep redundant copies of each element of not only \dir[j] but also
\dir[j-1].
The local memory overhead of \(2n / \nn\) vector elements per node and \(2n\)
vector elements in total is negligible compared to the overall memory
requirement \(\O{n^2}\) of the PCG solver (cf.\ Sec.~\ref{sec:background:pcg}).

The procedure to reconstruct the complete state of the PCG solver is outlined
in Alg.~\ref{alg:esr:pcg:2}.
It is assumed that a preconditioner \(\matP \coloneqq \Minv\) is given.
Variants for cases where \M (not \Minv) or a split preconditioner \(\M = \matL
\LT\) is given are shown in \cite[Alg.~3 and~5]{Pachajoa2018a}.
The scalar \(\beta^{(j-1)}\) can easily be recovered since it is replicated on
every node (cf.\ Sec.~\ref{sec:background:pcg}), i.e., \(\beta^{(j-1)}\) has
the same value on all nodes.
Furthermore, the reconstruction of the lost parts \res[j][\f] and \z[j][\f] of
the (preconditioned) residuals \res[j] and \z[j] takes place on the replacement
node \f.
Matrix \(\A_{\I[\f], \I[\f]}\) of the linear system in
line~\ref{alg:esr:pcg:2:solve} of Alg.~\ref{alg:esr:pcg:2} is SPD, has full
rank, and is much smaller than the full system matrix \A.
Therefore, this linear system can be solved locally on the replacement node \f
(only the involved vectors need to be gathered first on node \f).
Detailed derivations of the ESR approach can be found in \cite{chen2011esr,
Pachajoa2018a}.

\section{Single node failure}
\label{sec:single}

We now discuss details of how and where to store the redundant copies of the
two most recent search directions.
For this purpose, we review the strategy proposed by Chen \cite{chen2011esr}
for protecting the PCG method against a single node failure.
However, as we will see, this strategy is not suitable for multiple
simultaneous or overlapping node failures.
Later, in Sec.~\ref{sec:multiple}, we present our new strategy for handling
multiple simultaneous or overlapping node failures.

In PCG, the SpMV operation for obtaining \(\vec[j]{u} = \A \dir[j]\), which can
be rewritten as
\begin{equation}
\label{eq:single:chen_matrix_partition}
\begin{aligned}
\vec[j][1]{u} &= \A_{\I[1], \I[1]} \dir[j][1] +
                 \A_{\I[1], \I[2]} \dir[j][2] +
                 \dots +
                 \A_{\I[1], \I[\nn]} \dir[j][\nn] \\
\vec[j][2]{u} &= \A_{\I[2], \I[1]} \dir[j][1] +
                 \A_{\I[2], \I[2]} \dir[j][2] +
                 \dots +
                 \A_{\I[2], \I[\nn]} \dir[j][\nn] \\
&\mathrel{\makebox[\widthof{=}]{\vdots}} \\
\vec[j][\nn]{u} &= \A_{\I[\nn], \I[1]} \dir[j][1] +
                   \A_{\I[\nn], \I[2]} \dir[j][2] +
                   \dots +
                   \A_{\I[\nn], \I[\nn]} \dir[j][\nn],
\end{aligned}
\end{equation}
is performed at each iteration (cf.\ lines~\ref{alg:pcg:alpha}
and~\ref{alg:pcg:r} of Alg.~\ref{alg:pcg}).
When ignoring possible optimizations due to the sparsity pattern of \A,
\dir[j][i] is sent from node \(i\) to node \(k\) in iteration \(j\) such that
\(\vec[j][k]{u} = \A_{\I[k], \I[1]} \dir[j][1] + \dots + \A_{\I[k], \I[i]}
\dir[j][i] + \dots + \A_{\I[k], \I[\nn]} \dir[j][\nn]\) can be computed on node
\(k\).
For a more optimized algorithm that, during SpMV, only sends the minimum set of
elements required due to the sparsity pattern of \A, we define
\begin{equation}
\label{eq:single:chen_sets}
\begin{aligned}
S_i \coloneqq~&\text{all elements of}~\dir[j][i], \\
S_{ik} \coloneqq~&\text{elements of}~\dir[j][i]~\text{sent to
  node}~k~\text{computing}~\A \dir[j], \\
R_i \coloneqq~&\!\left(\bigcup_{k=1}^{i-1} S_{ik}\right) \cup
                 \left(\bigcup_{k=i+1}^{\nn} S_{ik}\right)\!,\,
\text{and}~R_i^c \coloneqq S_i - R_i
\end{aligned}
\end{equation}
in accordance with Chen \cite{chen2011esr}.
\dir[j][i] can be completely recovered after a node failure if \(R_i = S_i\).
In order to ensure this, Chen proposes to send \(R_i^c\) to node \(d_i
\coloneqq (i+1) \bmod \nn\) (together with \(S_{i d_i}\)).

Unfortunately, this strategy is not capable of coping with simultaneous or
overlapping failures of multiple nodes.
For example, if both nodes~\(i\) and~\(i+1\) fail simultaneously and
\(R_i^c \neq \emptyset\), the search direction vector elements in \(R_i^c\)
are lost and the state of the PCG solver cannot be reconstructed.
The problem obviously worsens if more than two nodes fail simultaneously.

\section{Multiple node failures}
\label{sec:multiple}

In this section, we extend the ESR approach for coping with multiple
simultaneous or overlapping node failures.
Originally, the ESR method considers exactly one node failure at a time, i.e.,
it is assumed that the reconstruction process finishes before another node
failure occurs (cf.\ Sec.~\ref{sec:background:esr} and Sec.~\ref{sec:single}).
For coping with up to \(\nredu < \nn\) uniformly distributed node failures that
may overlap in time, we need to keep \nredu redundant copies of each block of
the two most recent search directions \dir[j-1] and \dir[j] on \nredu different
compute nodes \cite{Pachajoa2018a}.
In the following, we design a resilient algorithm based on this idea in detail
and analyze its communication overhead.

\subsection{Tolerating multiple node failures}
\label{sec:multiple:algorithm}

To keep \nredu redundant copies of each block of the two most recent search
direction vectors, a similar strategy can be pursued as in the special case
\(\nredu = 1\).
Let \((S_i, m_i)\) be a multiset with the multiplicity
\begin{equation}
\label{eq:multiple:multiplicity}
\begin{aligned}
m_i \colon S_i \to~&\N_0 \\
s \mapsto~&\text{number of nodes}~s~\text{is sent to} \\[-1ex]
          &\text{during the computation of}~\A \dir[j].
\end{aligned}
\end{equation}
Note that we assume here---as it is common for SpMV---that \(s\) is only sent
to nodes with corresponding non-zero entries in their rows of \A.
Hence, the number of nodes \(s\) is sent to depends on the sparsity pattern of
\A.
Comparing the definition of the multiplicity \(m\) in
Eqn.~(\ref{eq:multiple:multiplicity}) with the definition of \(R_i^c\) in
Eqn.~(\ref{eq:single:chen_sets}), it follows that
\begin{equation}
R_i^c = \left\{ s \in S_i \; \middle| \; m_i(s) = 0 \right\}.
\end{equation}
For supporting up to \nredu simultaneous (or overlapping) node failures, we
need to store at least \nredu redundant copies of each element of \dir[j][i],
\(i \in \{1, 2, \dots, \nn\}\), on \nredu different nodes other than node
\(i\).
Let
\begin{equation}
\label{eq:multiple:rec}
d_{ik} \coloneqq \left\{
\begin{array}{ll}
\left( i + \left\lceil \frac{k}{2} \right\rceil \right) \bmod \nn, &
  \text{if}~k~\text{odd} \\
\left( i - \frac{k}{2} \right) \bmod \nn, & \text{if}~k~\text{even}
\end{array}
\right.
\end{equation}
and let \(g_i(s)\) denote the number of sets \(S_{i d_{ik}}\) with \(s \in
S_{i d_{ik}}\) for all \(k \in \{1, 2, \dots, \nredu\}\).
Then, the required redundancy for tolerating up to \nredu simultaneous node
failures can be achieved by sending
\begin{equation}
\label{eq:multiple:vulnerable_entries}
R_{ik}^c \coloneqq
  \left\{ s \in S_i \; \middle| \; s \notin S_{i d_{ik}} \; \land \;
                                   m_i(s) - g_i(s) \leq \nredu - k \right\}
\end{equation}
to node \(d_{ik}\) for all \(i \in \{1, 2, \dots, \nn\}\) and \(k \in \{1, 2,
\dots, \nredu\}\).
Note that the sets \(R_{ik}^c\) are of minimal size such that the required
number \nredu of redundant copies of each search direction vector element is
ensured.
It holds that
\(|R_{i 1}^c| \geq |R_{i 2}^c| \geq \dots \geq |R_{i \nredu}^c|\).

The strategy proposed in Eqn.~(\ref{eq:multiple:rec}) for selecting the nodes
to keep the redundant copies of \dir[j-1][i] and \dir[j][i] is a reasonably
good heuristic for minimizing communication overheads during SpMV if we assume
that the entries of the system matrix \A are mostly clustered around the
diagonal (since it then is likely that there are some elements which have to be
sent anyway from node \(i\) to node \(d_{ik}\) and, thus, there is no extra
latency for establishing a new connection; see Sec.~\ref{sec:sparsity} for a
more detailed discussion).
For matrices with very different sparsity patterns, strategies different from
Eqn.~(\ref{eq:multiple:rec}) may be preferable.
A comprehensive analysis of the interaction between a given sparsity pattern
and the optimal choice of the ``backup nodes'' is work in progress.

When the backup strategy based on Eqns.~(\ref{eq:multiple:rec})
and~(\ref{eq:multiple:vulnerable_entries}) is employed, we have \(\nredu + 1\)
copies of each element of \dir[j-1][i] and \dir[j][i] on \(\nredu + 1\)
different nodes (including node \(i\) that owns the block) and the two most
recent search directions \dir[j-1] and \dir[j] can be fully recovered after a
simultaneous or overlapping failure of up to \nredu arbitrary nodes.
If the node failures do not happen simultaneously but are overlapping in time,
i.e., more node failures occur during the reconstruction phase, the
reconstruction process must be restarted after each node failure (an efficient
implementation can of course skip steps that have already been performed and
are not affected by the subsequent node failures).

We consider \(\nfail \leq \nredu\) node failures.
Let \(\f_1, \f_2, \dots, \f_{\nfail}\) denote the nodes that fail.
Then, we can define \(\I[\f] \coloneqq \I[\f_1] \cup \I[\f_2] \cup \dots \cup
\I[\f_{\nfail}]\) and use a similar reconstruction procedure as in the case of
a single node failure.
Some of the reconstruction steps of Alg.~\ref{alg:esr:pcg:2} can be performed
locally on each of the replacement nodes \(\f_1, \f_2, \dots, \f_{\nfail}\).
However, for computing the matrix-vector products and solving the linear
systems in lines~\ref{alg:esr:pcg:2:pc_rhs} to~\ref{alg:esr:pcg:2:solve} of
Alg.~\ref{alg:esr:pcg:2}, additional communication between the \nfail
replacement nodes is necessary.
In Sec.~\ref{sec:multiple:analysis}, we show analytically that the capability
of tolerating up to \nredu node failures may incur increased communication
cost.
Hence, \nredu should be chosen only as large as necessary for handling the
expected number of simultaneous or overlapping node failures on a given
parallel computer.

\subsection{Analysis}
\label{sec:multiple:analysis}

Communication time usually is the main cost for a parallel algorithm,
dominating the computation time \cite{Ballard2014a}.
For analyzing the communication overhead of sending the additional elements of the
sets \(R_{ik}^c\) as defined in Eqn.~(\ref{eq:multiple:vulnerable_entries}), we adopt
a latency-bandwidth communication model \cite{Chan2007a} and assume that the
communication cost solely depends on latencies \(\latency_{ik} > 0\)---which may vary
for different sending nodes \(i\) and corresponding receiving nodes \(d_{ik}\)
according to Eqn.~(\ref{eq:multiple:rec})---and cost \(\invbw > 0\) per vector
element.
We further assume that each node is able to send and receive exactly one
element at a time.
If \(S_{i d_{ik}} \neq \emptyset\), the additional elements of \(R_{ik}^c\) are
sent together with the elements of \(S_{i d_{ik}}\), which have to be sent
anyway during computing the sparse matrix-vector product \(\A \dir[j]\).

If this is the case for all pairs of nodes for a fixed \(k \in \{1, 2, \dots,
\nredu\}\), which we refer to as communication round \(k\), no extra latency
cost applies in that round, and the overhead is \(\max_i |R_{ik}^c| \invbw\).
In contrast, if \(\forall i \in \{1,2,\dots,\nn\} \colon S_{i d_{ik}} =
\emptyset\) in communication round \(k \in \{1,2,\dots,\nredu\}\), extra
latencies \(\latency_{ik}\) incur for all pairs of nodes, and the overhead is
\(\max_i (\latency_{ik} + |R_{ik}^c| \invbw) \leq \max_i \latency_{ik} +
\max_i |R_{ik}^c| \invbw\).
Hence, in communication round \(k \in \{1,2,\dots,\nredu\}\), it holds for the
communication overhead \CommOverhead that
\begin{align*}
0 &\leq \max_i \left|R_{ik}^c\right| \invbw
   \leq \CommOverhead \\
  &\leq \max_i \left(\latency_{ik} + \left|R_{ik}^c\right| \invbw\right)
   \leq \max_i \latency_{ik} + \max_i \left|R_{ik}^c\right| \invbw,
\end{align*}
where \(i \in \{1,2,\dots,\nn\}\) and \(\max_i |R_{ik}^c| \invbw = 0\) if and
only if \(\forall i \colon |R_{ik}^c| = 0\), i.e., there are at least \(\nredu
- k + 1\) redundant copies of all elements of \dir[j] due to the sparsity
pattern of \A.
For all \nredu communication rounds, it follows that
\begin{align*}
0 &\leq \max_i \sum_{k=1}^{\nredu} \left|R_{ik}^c\right| \invbw
   \leq \sum_{k=1}^{\nredu} \max_i \left|R_{ik}^c\right| \invbw
   \leq \CommOverhead \\
  &\leq \sum_{k=1}^{\nredu} \max_i \left(\latency_{ik} + \left|R_{ik}^c\right|
        \invbw\right)
   \leq \sum_{k=1}^{\nredu} \left(\max_i \latency_{ik} + \max_i
        \left|R_{ik}^c\right| \invbw\right) \\
  &= \sum_{k=1}^{\nredu} \underbrace{\max_i \latency_{ik}}_{\leq \,
     \latency_{\max}} +
     \sum_{k=1}^{\nredu} \underbrace{\max_i \left|R_{ik}^c\right|}_{\leq \,
     \left\lceil \frac{n}{\nn} \right\rceil} \invbw
   \leq \nredu \latency_{\max} + \nredu \left\lceil \frac{n}{\nn} \right\rceil
        \invbw,
\end{align*}
where \(\latency_{\max} \coloneqq \max_{i,k} \latency_{ik}\).
Hence, the communication overhead \CommOverhead for keeping \nredu redundant copies
of all elements of \dir[j] lies between the lower bound 0 and the upper bound
\(\nredu \left( \latency_{\max} + \left\lceil \frac{n}{\nn} \right\rceil \invbw
\right)\).
The actual communication overhead within that interval entirely depends on the
sparsity pattern of \A, which determines the elements in \(R_{ik}^c\).

\section{Influence of the sparsity pattern}
\label{sec:sparsity}

In this section, we take a closer look at the implications of our theoretical
analysis in Sec.~\ref{sec:multiple}.
As we showed in Sec.~\ref{sec:multiple:analysis}, there exist matrices with
sparsity patterns which lead to \emph{zero} communication overhead during the
failure-free execution of the PCG solver, i.e., no extra communication is
necessary for distributing \nredu redundant copies of all elements of the search
direction vector \dir[j] during the computation of the sparse matrix-vector
product \(\A \dir[j]\).
On the other hand, other matrix sparsity patterns may lead to significant
communication overhead during the SpMV operation.

Independent of the particular strategy for selecting the backup nodes, it is
clearly beneficial for having a low communication overhead if \(m_i(s) \geq
\nredu\) (cf.\ Eqn.~(\ref{eq:multiple:multiplicity})) holds for most or even
all \(s \in S_i\) and \(i \in \{1, 2, \dots, \nn\}\) or, in other words, if
most or all elements of \dir[j] anyway---i.e., due to the sparsity pattern of
\A---have to be communicated to at least \nredu different nodes during the
computation of the product \(\A \dir[j]\).

If this is not the case, a possibly considerable number of extra vector
elements has to be transferred during the SpMV operation.
Since the number of extra elements to be sent is determined by the sparsity pattern
of \A, communication cost can then only be reduced by avoiding extra latencies,
i.e., by sending the extra elements to nodes where also other elements have to
be sent to.
In general, our strategy defined by Eqns.~(\ref{eq:multiple:rec})
and~(\ref{eq:multiple:vulnerable_entries}) performs well if \A is not
\emph{too} sparse within a bandwidth of \(\ceil{\nredu n / (2 \nn)}\) around
the diagonal (but can still be very sparse overall).
More formally, if for all \(i \in \{1, 2, \dots, \nn\}\) and \(k \in \{1, 2,
\dots, \nredu\}\) at least one element in each submatrix \(\A_{\I[d_{ik}],
\I[i]}\) is not equal to zero, no cost because of extra latencies is incurred.
The corresponding proofs are straightforward, but we are omitting them due to space
restrictions.

\section{Implementation}
\label{sec:implementation}

Up to this point,
we have analyzed the algorithms in theoretical terms.
We now describe how the reconstruction is realized in a real-life, finite-precision machine.

We implement Chen's algorithm \cite{chen2011esr} and our novel extensions as described in Sec.~\ref{sec:single} and
Sec.~\ref{sec:multiple} using the PETSc framework
\cite{Balay1997a,petsc-user-ref}.
PETSc provides the CG solver and linear algebra operations,
and it also manages communication between nodes.
We use operations offered by the framework to transfer the information required for the recovery from node failures.
We use a block Jacobi as a preconditioner during the regular operation of the solver,
solving the preconditioner blocks exactly.

To impart fault tolerance,
the SpMV is modified to transfer the additional data required to obtain the desired level of data redundancy.
In PETSc, the SpMV operation is realized with a generalized scatter:
A node determines, from the non-zero entries in its matrix rows,
what vector components it requires from its neighbors to perform the SpMV product.
With this information,
PETSc collectively creates a \emph{communication context} for the generalized scatter operation,
defining what entries of a distributed vector are communicated,
and where they must be communicated to.

In our experiments, we simulate node failures.
Instead of taking down nodes and producing replacements during the reconstruction phase,
a node will perform the operations and communication required to restore the solver state.
The reconstruction process requires sending the surviving entries of the search direction vector to replacement nodes.
In our experiments, this is achieved by reversing the communication that takes place during the matrix-vector product.
PETSc already provides this functionality.
However, reversing the communication that occurs during the matrix-vector product is not a well-defined operation.
To see this, imagine a communication context that dictates that the entry in position \(i_0\) of a vector, located in some node \(A\),
must be transferred to positions \(i_1\), in node \(B\), and \(i_2\), in node \(C\).
In the reverse communication process,
entries in positions \(i_1\) and \(i_2\) will hold candidates for the value of that entry to be transferred to position \(i_0\).
In the absence of node-failures, both candidates will be the same value,
because they are copies of the original entry in \(i_0\),
and the operation is then well defined,
but, in the event of a node failure,
it is possible that one of these entries is lost and the resulting candidates are different, conflicting values.
Therefore, in such cases, the reverse communication process used in the reconstruction could be non-deterministic.
We cope with this issue by keeping the search directions in the nodes simulating a node failure.
If this information is stored, communication with the reversed context is deterministic,
because there would never be a conflict between the candidates.

This problem arises because we use the reverse of a communication context intended for SpMV.
In a more optimized implementation,
a tailored communication context can be produced
after the node failure takes place,
which avoids this problem altogether by selecting the entries that we need for the reconstruction.

In our implementation, the linear system arising in line~\ref{alg:esr:pcg:2:solve} of Alg.~\ref{alg:esr:pcg:2},
is solved using a PCG solver assembled with global operations.
In particular, the matrix-vector products of the submatrix \(\A_{\I[\f], \I[\f]}\) and the subvector \x[j][\f]
are performed by multiplying the entire matrix \A with a modified vector \x[j],
whose appropriate entries were set to zero.
The desired  \(\A_{\I[\f], \I[\f]} \x[j][\f]\) is a subvector of the result of this global operation.
This is less efficient than working with the actual submatrix of \A,
but some of the changes we introduced to increase redundancy conflict with PETSc' ability to create submatrices.
The cost to reconstruct the solution, however, remains very small compared to the overall runtime.
The CG solver for the subsystem uses a block Jacobi preconditioner,
with blocks matching the process' index set.
We use an approximate solver based on ILU factorization for the blocks.

\subsection*{Avoiding loss of orthogonality}
\label{sec:loss_of_orthogonality}

For this section it is useful to distinguish between the \emph{solver residual}, that is,
vector \res[j] from Alg.~\ref{alg:pcg},
and the vector \(\B - A \x[j] \).
These vectors are, in general, not equal in a finite-precision machine.

The CG algorithm in floating-point arithmetic undergoes loss of orthogonality,
where roundoff error accumulates and the conjugacy of the search directions is lost as the solver progresses.
Consequently, in regular PCG the solver residual
and the vector \(\B-\A \x\) will differ slightly after convergence.
Because we work with finite precision,
and because we solve the local linear system in the reconstruction process
(line \ref{alg:esr:pcg:2:solve} of Alg. \ref{alg:esr:pcg:2}) iteratively,
our algorithm only reconstructs an approximation of the solver state before the node failures take place,
thus potentially further contributing to this loss of orthogonality:
Consequently, the ESR solver residual after convergence can be larger
than the solver residual of PCG.
The largest source of deviations is the solution of the local linear system of line~\ref{alg:esr:pcg:2:solve} of Alg.~\ref{alg:esr:pcg:2}.
The loss of orthogonality relative to regular PCG can be controlled with the tolerance of this linear system.

To compare the accuracies of ESR and PCG in this regard,
we define the \emph{relative residual difference} metric:

\begin{align}
\begin{split}
\relresdiff{\mbox{\tiny ESR}} &= \frac{\|\res_{\mbox{\tiny ESR}}\|_2 - \|\B-\A \x_{\mbox{\tiny ESR}}\|_2}{\|\B-\A \x_{\mbox{\tiny ESR}}\|_2}\\
\relresdiff{\mbox{\tiny PCG}} &= \frac{\|\res_{\mbox{\tiny PCG}}\|_2 - \|\B-\A \x_{\mbox{\tiny PCG}}\|_2}{\|\B-\A \x_{\mbox{\tiny PCG}}\|_2}.
\end{split}
\label{eq:sanity_metric}
\end{align}

Here, \(\res_{\mbox{\tiny ESR}}\) and \(\res_{\mbox{\tiny PCG}}\) are the solver residual vectors of ESR with reconstruction and reference PCG respectively after convergence,
and \(\x_{\mbox{\tiny ESR}}\) and \(\x_{\mbox{\tiny PCG}}\) are the corresponding iterands.

A side-by-side comparison of the ESR method and regular PCG can show that the former is as accurate as the latter.
In Sec.~\ref{sec:experiments:results} we show that the effects of using finite-precision arithmetic can be made negligible.
Since node failures are uncommon and reconstruction is relatively cheap to perform,
we can set the tolerance for the local system to a very small value,
so that ESR converges,
while the reconstruction overhead remains low.

\section{Numerical experiments}
\label{sec:experiments}

Now we summarize our experimental setup and discuss our results.

\subsection{Experimental setup}
\label{sec:experiments:setup}

We use SPD matrices from the SuiteSparse Matrix Collection \cite{Davis2011a} as test problems,
selecting problems from different application areas.
Their properties are summarized in Table \ref{tab:testproblems}.
We select medium (\parabolicfem, \offshore) and large (\gthree-\audikw) size problems.
The latter are among the largest available SPD matrices in the SuiteSparse Matrix Collection.
There are problems with different numbers of non-zeros.
Large matrices with relatively few non-zeros are common,
but problems with more non-zero entries are more expensive to compute and would benefit the most from resilience schemes to protect the resource investment.

\begin{table}[h!]
\caption{SPD matrices from \cite{Davis2011a} used in the experiments.
\(n\): Problem size.
\(NNZ\): Number of non-zero entries.
The matrices are ordered by increasing number of non-zero entries,
with a larger ID indicating a larger number of non-zeros.
}
\begin{center}
\setlength{\tabcolsep}{0.4em}
\begin{tabular}{lclrr}
\toprule
\textbf{Name} & \textbf{Id} & \textbf{Problem type} & \textbf{\(n\)} & \textbf{\(NNZ\)} \\
\midrule
parabolic\_fem   & \parabolicfem   & Fluid dynamics       &   525 825     &  3 674 625 \\
offshore         & \offshore       & Electromagnetics     &   259 789     &  4 242 673 \\
G3\_circuit      & \gthree         & Circuit simulation   & 1 585 478     &  7 660 826 \\
thermal2         & \thermal        & Thermal              & 1 228 045     &  8 580 313 \\
Emilia\_923      & \emilia         & Structural           &   923 136     & 40 373 538 \\
Geo\_1438        & \geo            & Structural           & 1 437 960     & 60 236 322 \\
Serena           & \serena         & Structural           & 1 391 349     & 64 131 971 \\
audikw\_1        & \audikw         & Structural           &   943 695     & 77 651 847 \\
\bottomrule
\end{tabular}
\end{center}
\label{tab:testproblems}
\end{table}

Our experiments are run on 128 nodes of the VSC3 system of the Vienna Scientific Cluster.
Although our algorithm is well suited for multiple processes per node, we use
only one process per node in our experiments.
The argument for this decision is twofold:
Firstly, from the point of view of resilience, the number of processes per
node (cf.\ Sec.~\ref{sec:introduction:setting}) makes no difference since the
redundant vector elements always have to be stored on a different node (not
just in the memory of another process).
Secondly, the runtimes obtained in our experiments are based on simulations of
node failures without ULFM (cf.\ Secs.~\ref{sec:introduction:setting}
and~\ref{sec:implementation}).
Hence, the relative runtime differences are more significant than the absolute
runtimes (and, thus, improvements of the absolute runtimes due to multiple
processes per node).
Experiments for a matrix are run on the same set of nodes of VSC3.
The system's topology is a fat tree.
We use the following libraries:
\begin{itemize*}
\item Intel MPI 5.1.3
\item PETSc 3.10.4
\item Intel MKL 2018.3
\end{itemize*}.
We use the Intel C compiler 18.0.5 with compiler flags \texttt{-O3},
\texttt{-march=native} and \texttt{-mtune=native}.
We terminate the solver once the relative residual norm has been reduced by a factor of \(10^{8}\).
The local linear system used during the reconstruction is terminated once its residual norm is reduced by a factor of \(10^{14}\).

We measure the runtime of the solver for several problem settings.
Node failures are introduced once for each simulation,
with either one, three or eight simultaneous node failures taking place
at either 20\%, 50\% or 80\% of the solver progress (measured in number of iterations).
These failures are placed in contiguous ranks.
Simultaneous node failures can well be caused by a faulty switch,
therefore it seems like a realistic assumption that they are clustered:
The node failures are introduced in neighbouring ranks either at the beginning or in the center of the vector, starting from rank 0 or 64 respectively.
Additionally, we have results for runs without failures with either one, three or eight redundant copies.
Each measurement in this test constellation is repeated at least 5 times.

\subsection{Experimental results}
\label{sec:experiments:results}
Our experimental results are summarized in Tab.~\ref{tab:summary_table}.
Statistics for node failures are computed over at least 15 values:
at least 5 measurements for node failures introduced at either 20\%, 50\% and 80\% of the solver's progress.

As expected, a larger number of redundant copies leads to larger overheads.
In general, the reconstruction time remains small,
with larger relative reconstruction costs for matrices with smaller runtimes,
where the absolute cost is consequently smaller.

The overall relative overhead with node failures, shown in the last three columns of Tab.~\ref{tab:summary_table},
corresponds to the sum of the relative overhead of the undisturbed case
plus the relative runtime cost of the reconstruction.
These columns roughly match the sum, as expected,
with some variation arising from the variation in runtime of running the code in a real machine,
and from differences in the number of iterations caused by the reconstruction (cf.\ \cite{Pachajoa2018a}).

Tab. \ref{tab:summary_table} also shows that the location of the node failure does affect the reconstruction cost,
as the runtime is, in general, different for node failures at the start (lower indexes) or the center (middle indexes) of the vectors.
These differences come from different diagonal linear systems in the reconstruction:
Submatrices formed from the index sets of different failed nodes have different properties, and they will not converge at the same rate with an ILU preconditioner,
thus affecting the reconstruction time.

\begin{table*}
\begin{center}
\caption{Summarized experimental results.
Values are aggregated for experiments with different times (iteration numbers) when the failures are introduced.
Matrices are ordered by descending reference runtime,
with the ID number of a matrix growing with increasing number of non-zeros.
\(t_0\): average reference time for the reference runs with regular (non-fault tolerant) PCG.
\nredu: number of redundant copies for the search direction.
\nfail: number of simultaneous node failures introduced.
The column ``relative overhead undisturbed'' shows the mean overhead of modified ESR-capable PCG,
with a given number of redundant copies \(\nredu\),
with respect to the reference time \(t_0\) if no reconstruction takes place.
The columns marked ``relative reconstruction time'' indicate the time that it takes to reconstruct the state of the solver,
expressed as a percentage of the reference time \(t_0\),
plus/minus a standard deviation.
The columns marked ``relative overhead with failures'' show the mean, plus/minus a standard deviation, of the overall relative overhead,
that is, the relative overhead for the total time until convergence,
when node failures occur and the state of the solver is reconstructed, with respect to the reference runtime \(t_0\).
}
\label{tab:summary_table}
\begin{tabular}{
l
S[table-format=1.2,table-number-alignment=left]
*{3}{S[table-format=1.1,table-number-alignment=left]}
l
l
*{3}{S[table-format=2.1(1),table-number-alignment=left]}
l
*{3}{S[table-format=2.1(1),table-number-alignment=left]}
}

\toprule

\multirow{2}{*}{\textbf{ID}} &
{\multirow{2}{*}{\textbf{\(t_0\) [s]}}} &
\multicolumn{3}{c}{\textbf{\makecell{Relative overhead \\ undisturbed [\%]}}} &
{\multirow{2}{*}{\textbf{\makecell{\makecell{Failure \\ location}}}}} &
&
\multicolumn{3}{c}{\textbf{\makecell{Relative reconstruction \\ time [\%]}}} &
&
\multicolumn{3}{c}{\textbf{Overhead with failures [\%]}} \\
&
&
\textbf{\(\nredu = 1\)} &
\textbf{\(\nredu = 3\)} &
\textbf{\(\nredu = 8\)} &
&
&
\textbf{\(\nfail = \nredu = 1\)} &
\textbf{\(\nfail = \nredu = 3\)} &
\textbf{\(\nfail = \nredu = 8\)} &
&
\textbf{\(\nfail = \nredu = 1\)} &
\textbf{\(\nfail = \nredu = 3\)} &
\textbf{\(\nfail = \nredu = 8\)} \\
\midrule
\emilia      &  43.29 &  0.4  &  5.1 &  16.3   & start  & &  0.2\pm 0.1 &  0.3\pm 0.1 &  0.6\pm 0.1 & &   1.2\pm 0.5 &   6.0\pm 0.3 &  19.6\pm 0.7 \\
             &        &       &      &         & center & &  0.0\pm 0.1 &  0.1\pm 0.1 &  0.4\pm 0.1 & &   1.2\pm 0.6 &   5.6\pm 0.3 &  20.5\pm 6.0 \\
\rule{0pt}{3ex}
\audikw      &  19.12 &  1.5  &  2.2 &   8.2   & start  & &  1.4\pm 0.1 &  3.5\pm 0.2 & 10.7\pm 0.2 & &   3.8\pm 1.0 &   5.6\pm 0.7 &  19.9\pm 0.9 \\
             &        &       &      &         & center & &  0.4\pm 0.1 &  0.8\pm 0.1 &  1.4\pm 0.1 & &   1.9\pm 0.4 &   2.8\pm 1.0 &  10.4\pm 1.1 \\
\rule{0pt}{3ex}
\geo         &  11.70 &  0.3  &  3.1 &  15.1   & start  & &  1.3\pm 0.1 &  2.4\pm 0.2 &  5.3\pm 0.2 & &   1.4\pm 0.3 &   5.9\pm 0.4 &  21.2\pm 1.2 \\
             &        &       &      &         & center & &  0.3\pm 0.1 &  0.6\pm 0.1 &  1.6\pm 0.1 & &   0.7\pm 0.4 &   4.0\pm 0.3 &  18.3\pm 0.7 \\
\rule{0pt}{3ex}
\serena      &   6.48 &  0.2  &  4.3 &  13.6   & start  & &  2.6\pm 0.1 &  8.2\pm 0.3 & 23.4\pm 0.6 & &   2.9\pm 0.3 &  13.7\pm 0.6 &  39.6\pm 1.4 \\
             &        &       &      &         & center & &  1.2\pm 0.1 &  1.5\pm 0.1 & 21.1\pm 0.7 & &   1.2\pm 0.2 &   7.1\pm 0.6 &  36.7\pm 1.7 \\
\rule{0pt}{3ex}
\thermal     &   6.31 &  8.2  & 22.8 &  65.6   & start  & &  2.0\pm 0.1 &  3.1\pm 0.1 &  4.9\pm 0.3 & &   9.6\pm 0.8 &  26.4\pm 1.1 &  66.0\pm 7.9 \\
             &        &       &      &         & center & &  1.3\pm 0.1 &  3.8\pm 0.2 &  5.2\pm 0.4 & &   9.2\pm 2.6 &  35.7\pm 2.2 &  63.4\pm 9.3 \\
\rule{0pt}{3ex}
\parabolicfem&   2.25 &  3.6  &  5.1 &  24.5   & start  & &  0.3\pm 0.1 &  0.3\pm 0.1 &  0.4\pm 0.1 & &   5.1\pm 1.2 &   4.3\pm 0.9 &  23.8\pm 2.1 \\
             &        &       &      &         & center & &  0.3\pm 0.1 &  0.3\pm 0.1 &  0.4\pm 0.1 & &   5.0\pm 1.0 &   5.1\pm 1.3 &  25.4\pm 1.8 \\
\rule{0pt}{3ex}
\offshore    &   1.58 &  8.0  &  8.1 &  21.1   & start  & &  4.0\pm 0.2 &  7.0\pm 0.3 &  9.7\pm 0.4 & &   7.3\pm 4.6 &  15.6\pm 2.3 &  30.1\pm 5.5 \\
             &        &       &      &         & center & &  2.1\pm 0.3 &  3.5\pm 0.2 &  4.4\pm 0.3 & &  10.0\pm 4.6 &  14.2\pm 2.6 &  23.8\pm 3.0 \\
\rule{0pt}{3ex}
\gthree      &   1.16 &  5.0  & 24.1 &  91.3   & start  & &  0.6\pm 0.1 &  1.0\pm 0.1 &  1.9\pm 0.1 & &   6.3\pm 1.7 &  24.8\pm 1.8 &  93.6\pm 6.0 \\
             &        &       &      &         & center & &  8.0\pm 0.5 & 32.2\pm 1.2 & 58.1\pm 1.4 & &  14.8\pm 2.0 &  55.0\pm 3.2 & 147.6\pm 5.0 \\

\bottomrule

\end{tabular}

\end{center}
\end{table*}

Our experiments show that our algorithm is very efficient for matrices with many non-zeros contained in a band close to the diagonal.
Relatively denser matrices also take longer to reach convergence because of the longer time required for the matrix-vector product,
so it makes a lot of sense to protect the time investment in the solution process with a resilience technique like the one presented in this paper.

\begin{figure}
\begin{center}
\input{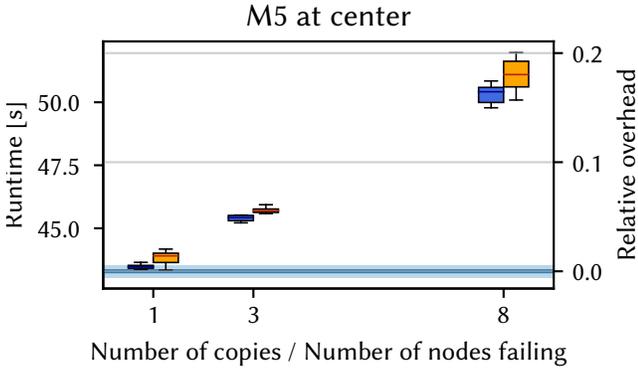}
\caption{
Runtimes and relative overhead for the matrix \emilia of Tab.~\ref{tab:testproblems},
for node failures introduced close to the center of the vector.
The blue line at the bottom of the figure represents the reference time obtained solving the system using PETSc without modifications, with respect to which the relative overhead is measured,
and the band around it extends for one standard deviation in each direction.
The \(x\)-axis indicates the number of copies the resilient solver holds.
Boxes include points in the interquartile range,
and whiskers extend up to 1.5 times the width of the interquartile range.
Blue boxes (to the left of a group) represent runs with the resilient solver without node failures.
Orange boxes (to the right of a group) represent runs with node failures.
In experiments with node failures, we introduce as many simultaneous failures as the solver can tolerate (one, three or eight failures respectively.)
Orange boxes include experiments for failures introduced at 20\%, 50\% or 80\% progress of the solver.
}
\label{fig:emilia_center}
\end{center}
\end{figure}

Fig.~\ref{fig:emilia_center} is an example of the runtimes and overheads obtained with our novel resilient algorithm
for the matrix \emilia.
For this matrix, the state reconstruction operation takes very little time:
The runtimes for cases with failures,
(orange boxes)
are very close to the failure-free cases
(blue boxes).
In this case,
the overhead for the method comes predominantly from the additional communication required to maintain redundant data.

\begin{figure}
\begin{center}
\input{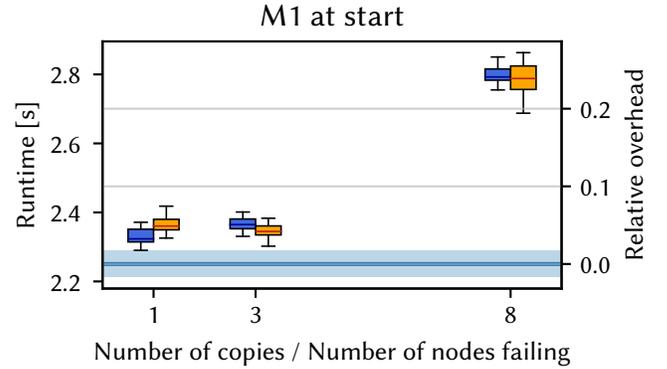}
\caption{
Runtimes  and relative overheads for matrix \parabolicfem with node failures occurring close to the start (lower indices) of the vectors.
This figure uses the same conventions as Fig.~\ref{fig:emilia_center},
and showcases a situation,
for three redundant copies,
where the solver converges faster after performing reconstruction after a node failure
due to the reduction of the number of iterations until convergence.
}
\label{fig:parabolic_start}
\end{center}
\end{figure}

In Fig.~\ref{fig:parabolic_start}
the boxes corresponding to three redundant copies
indicate a \emph{smaller} runtime for simulations with node failures than for the failure-free case.
As mentioned before,
this can happen,
since the number of iterations required after reconstruction can be a smaller than in the failure-free run
(cf. \cite{Pachajoa2018a}).

\begin{figure}
\begin{center}
\input{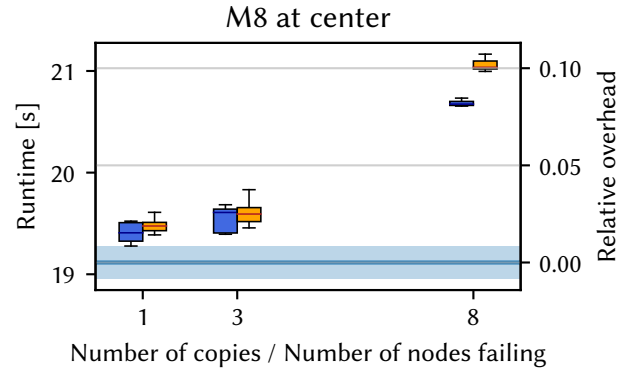}
\caption{
Runtimes and relative overheads for matrix \audikw with node failures occurring close to the start of the vectors.
This figure uses the same conventions as Fig.~\ref{fig:emilia_center},
and shows superlinear increase of the overhead with respect to the number of redundant copies held.
}
\label{fig:audikw_center}
\end{center}
\end{figure}

Fig.~\ref{fig:audikw_center} shows a test case where the overhead required to keep redundant data increases superlinearly with the number of node failures tolerated,
As explained in Sec.~\ref{sec:sparsity},
the growth of the overhead with the number of node failures tolerated strongly depends on the sparsity pattern of \A.
The matrix \audikw contains many non-zeros in a band around the diagonal in the middle indexes.
As expected from the analysis of Sec.~\ref{sec:sparsity},
this is a particularly favorable case for our method:
It can resist three simultaneous node failures with an overhead of around 2.5\%, and eight node failures with an overhead of around 10\%.

\begin{figure}
\begin{center}
\input{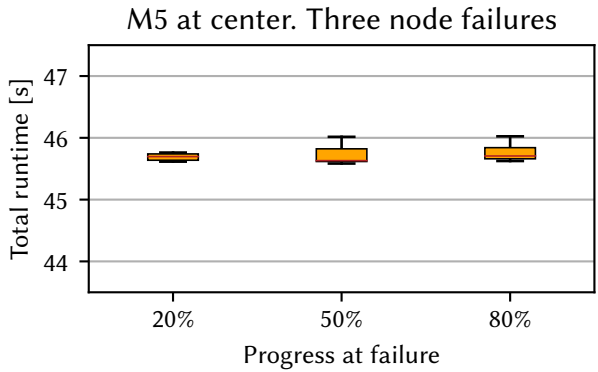}
\caption{
Runtime for matrix \emilia
when introducing the failure of three nodes
at the center (middle indexes) of the vectors
and at different iterations along the progress of the solver.
The boxes contain runtimes in the interquartile range,
and the whiskers extend up to 1.5 times the width of the interquartile range.
}
\label{fig:emilia_timestep}
\end{center}
\end{figure}

In general, the iteration at which the node failures are introduced has little influence on the runtime of the solver.
Fig.~\ref{fig:emilia_timestep} illustrates this for \emilia.
We observed the same behaviour for the other test cases.

In Sec.~\ref{sec:loss_of_orthogonality},
we discuss the potential loss of orthogonality that occurs when performing the reconstruction.
Tab.~\ref{table:precision} shows that the relative residual difference for our method is comparable to the one for the reference results,
even for its maximum value among all experiments for a given matrix.
The deviations for both methods are tiny in comparison to the reduction of the residual norm by a factor of \(10^8\) from the solver.

\begin{table}
\caption{Evaluation of the metric of Eqn.~(\ref{eq:sanity_metric}).
The first column shows the maximum value of the relative residual deviation for all experiments with node failures for a matrix.
The second column shows the relative residual deviation for the reference run.
}
\begin{center}
\begin{tabular}[t]{l
*{2}{S[table-format=-1.2e-1,table-number-alignment=left]}
}
\toprule

\textbf{ID} & \textbf{\makecell{max \relresdiff{\mbox{\tiny ESR}}}}& \textbf{\makecell{\relresdiff{\mbox{\tiny PCG}}}}\\

\midrule

\parabolicfem & 3.46e-7 & -1.63e-7 \\
\offshore     & 2.24e-7 &  1.86e-7 \\
\gthree       & 1.57e-7 &  1.57e-7 \\
\thermal      & 1.96e-7 &  9.06e-8 \\
\emilia       & 3.59e-5 &  1.81e-6 \\
\geo          & 8.80e-8 & -3.31e-8 \\
\serena       & 1.32e-7 & -7.24e-8 \\
\audikw       & 2.64e-3 &  1.49e-3 \\

\bottomrule
\end{tabular}

\end{center}
\label{table:precision}
\end{table}

\section{Conclusions}
\label{sec:conclusions}

In this paper, we first reviewed the ESR approach, which was initially proposed
by Chen~\cite{chen2011esr} and later refined by Pachajoa et
al.~\cite{Pachajoa2018a}, an efficient fault-tolerance technique to protect the
PCG method against a single node failure (cf.\ Secs.~\ref{sec:background:esr}
and~\ref{sec:single}).
We then proposed an enhancement to the ESR approach that allows the PCG method
to tolerate simultaneous or overlapping failures of \emph{multiple} nodes
(cf.\ Sec.~\ref{sec:multiple:algorithm}).
Our new strategy determines where to efficiently store redundant information in
order to support up to \(\nredu < \nn\) simultaneous node failures.
In a theoretical analysis, we found that the communication overhead due to the
distribution of the required additional redundant vector copies strongly
depends on the sparsity pattern of the given system matrix
(cf.\ Secs.~\ref{sec:multiple:analysis} and~\ref{sec:sparsity}).

In order to investigate the effects of floating-point arithmetic and the
runtime performance of our novel algorithm, we implemented it based on the
widely used library PETSc (cf.\ Sec.~\ref{sec:implementation}) and conducted
numerical experiments on 128 nodes of the Vienna Scientific Cluster
(cf.\ Sec.~\ref{sec:experiments}).
The results of our experiments with eight large sparse matrices from real-world
applications show that the proposed enhanced ESR approach is very efficient.
Compared to a non-resilient PCG run, we measured runtime overheads between
2.2\% and 24.1\% for an undisturbed run with up to three tolerated simultaneous
node failures and between 2.8\% and 55.0\% for a run with three actual
simultaneous node failures including reconstruction.

In future work, we are going to further enhance the ESR approach such that it
automatically adapts to different sparsity patterns of matrices.
Moreover, we want to investigate communication-avoiding PCG methods.
Another interesting future direction is to find a variant of the ESR approach
that does not depend on the availability of replacement nodes for failed nodes.

\begin{acks}
This work has been funded by the \grantsponsor{wwtf}{Vienna Science and
Technology Fund (WWTF)}{https://www.wwtf.at/} through project
\grantnum{wwtf}{ICT15-113}.
The computational results presented have been achieved using the Vienna
Scientific Cluster (VSC).
\end{acks}

\balance
\bibliographystyle{ACM-Reference-Format}
\bibliography{Bibliography}

\end{document}